\documentstyle[array,multicol,aps,rmp,psfig]{revtex}

\begin{document}

\title{\Large \bf Optimal sizes of dendritic and axonal arbors in a
topographic projection}
\author{Dmitri B. Chklovskii}
\address{Sloan Center for Theoretical Neurobiology, The Salk Institute, La Jolla, CA 92037}

\date{\today}
\draft
\tighten
\preprint{Submitted to {\it Nature Neuroscience}}
\maketitle

\begin{abstract}
I consider a topographic projection between two neuronal layers with
different densities of neurons. Given the number of output neurons
connected to each input neuron (divergence) and the number of input
neurons synapsing on each output neuron (convergence) I determine the
widths of axonal and dendritic arbors which minimize the total volume of
axons and dendrites. Analytical results for one-dimensional and
two-dimensional projections can be summarized qualitatively in the
following rule: neurons of the sparser layer should have arbors wider
than those of the denser layer. This agrees with the anatomical
data from retinal and cerebellar neurons whose morphology and
connectivity are known. The rule may be used to infer connectivity of
neurons from their morphology.
\end{abstract}

\begin{multicols}{2}
\section{INTRODUCTION}

Understanding brain function requires knowing connections between
neurons. However, experimental studies of inter-neuronal connectivity
are difficult and the connectivity data is scarce. At the same time
neuroanatomists possess much data on cellular morphology and have
powerful techniques to image neuronal shapes. This suggests using
morphological data to infer inter-neuronal connections. Such inference
must rely on rules which relate shapes of neurons to their connectivity.

\begin{figure}
\centerline{
\psfig{file=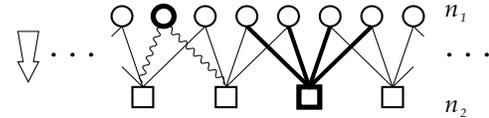,width=2.5in}
}
\vspace{0in} 
\setlength{\columnwidth}{3.4in}
\caption{Wiring diagram of a topographic projection between 
input (circles) and output (squares) layers of neurons. Divergence, 
$D$, is the number of outgoing connections (here, $D=2$) from an input neuron
(wavey lines). Convergence, $C$, is the number of connections incoming 
(here, $C=4$) to an output neuron (bold lines). Arrow shows the direction
of signal propagation.
\label{fig:cd}
}
\vspace{-0.1in}
\end{figure}

The purpose of this paper is to derive such a rule for a frequently
encountered feature in the brain organization: a topographic
projection. Two layers of neurons are said to form a topographic
projection if adjacent neurons of the input layer connect to adjacent
neurons of the output layer, Figure~\ref{fig:cd}. As a result, output
neurons form an orderly map of the input layer.

I characterize inter-neuronal connectivity for a topographic
projection by divergence and convergence factors defined as follows,
Fig.~\ref{fig:cd}. {\it Divergence}, $D$, of the projection is the
number of output neurons which receive connections from an input
neuron. {\it Convergence}, $C$, of the projection is the number of
input neurons which connect with an output neuron. I assume that these
numbers are the same for each neuron in a given layer. Furthermore,
each neuron makes the required connections with the nearest neurons of
the other layer. In most cases, this completely specifies the
wiring diagram.

A typical topographic wiring diagram shown in Fig.~\ref{fig:cd} misses an
important biological detail. In real brains, connections between cell
bodies are implemented by neuronal processes: axons carrying nerve
pulses away from the cell bodies and dendrites carrying signals
towards cell bodies.\cite{Cajal1} Therefore each connection is
interrupted by a synapse which separates an axon of one neuron from a
dendrite of another. Both axons and dendrites branch away from cell 
bodies forming arbors.

\begin{figure}
\centerline{
\psfig{file=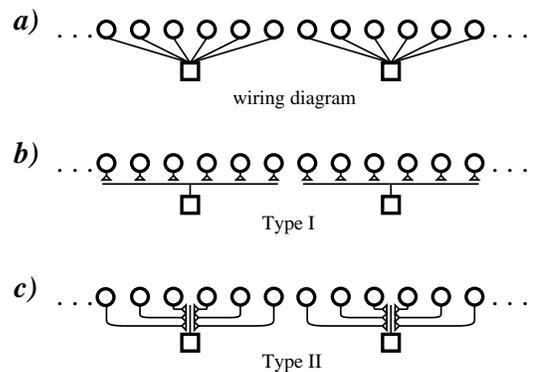,width=2.7in}
}
\vspace{0in} 
\setlength{\columnwidth}{3.4in}
\caption{Two different arrangements implement the
same wiring diagram. (a)Topographic wiring diagram with 
$C=6$ and $D=1$. (b)Arrangement with wide dendritic arbors and
no axonal arbors (Type I) (c) Arrangement with wide axonal arbors
and no dendritic arbors (Type II). Because convergence exceeds 
divergence type I has shorter wiring than type II.
\label{fig:menorah}
}
\vspace{-0.2in}
\end{figure}

In general, a topographic projection with given divergence and
convergence may be implemented by axonal and dendritic arbors of
different sizes, which depend on the locations of synapses.  For
example, consider a wiring diagram with $D=1$ and $C=6$,
Figure~\ref{fig:menorah}a. Narrow axonal arbors may synapse onto wide
dendritic arbors, Figure~\ref{fig:menorah}b or wide axonal arbors may
synapse onto narrow dendritic arbors, Figure~\ref{fig:menorah}c. 
I call these arrangements type I and type II, correspondingly. The 
question is: which arbor sizes are preferred?

I propose a rule which specifies the sizes of axonal arbors of input
neurons and dendritic arbors of output neurons in a topographic
projection: {\it High divergence/convergence ratio favors wide axonal
and narrow dendritic arbors while low divergence/convergence ratio
favors narrow axonal arbors and wide dendritic arbors.}
Alternatively, this rule may be formulated in terms of neuronal
densities in the two layers: {\it Sparser layer has wider arbors.}  In
the above example, divergence/convergence (and neuronal density) ratio
is 1/6 and, according to the rule, type I arrangement,
Figure~\ref{fig:menorah}b, is preferred.

In this paper I derive a quantitative version of this rule from the 
principle of wiring economy which can be summarized as follows.
\cite{Cajal2},\cite{Mitch1},\cite{Chern},\cite{Young},\cite{Cat} 
Space constraints require keeping 
the brain volume to a minimum. Because wiring (axons and dendrites) takes
up a significant fraction of the volume, evolution has probably
designed axonal and dendritic arbors in a way that minimizes their
total volume. Therefore we may understand the existing arbor sizes
as a result of wiring optimization.

To obtain the rule I formulate and solve a wiring optimization
problem. The goal is to find the sizes of axons and dendrites which
minimize the total volume of wiring in a topographic wiring diagram
for fixed locations of neurons. I specify the wiring diagram with 
divergence and convergence factors. Throughout
most of the paper I assume that the cross-sectional area of dendrites
and axons are constant and equal. Therefore, the problem reduces to
the wiring {\it length} minimization. My results are trivially
extended to the case of unequal fiber diameters as shown below.

Purves and coworkers\cite{PurHum},\cite{PurRub},\cite{PurLic} have 
previously reported empirical
observations which may be relevant to the present theory. They found
a correlation between convergence and complexity of dendritic arbors in 
sympathetic ganglia. Conclusive comparison of this data with the theory 
requires establishing topographic (or some other) wiring diagram and 
measuring axonal arbor sizes in this system.

In the next section I consider a one-dimensional version of the
problem. In this case, wirelength is minimized by wide dendritic and
no axonal arbors (Type I) for divergence less than convergence and by
no dendritic and wide axonal arbors (Type II) in the opposite case.
In Section \ref{sec:2d}, I consider a two-dimensional version of the
problem. If both convergence and divergence are much greater than one,
the optimal ratio of dendritic and axonal arbors equals to the square
root of convergence/divergence ratio.

I test the rule on the available anatomical data in Section \ref{sec:data}.
For several projections between retinal and cerebellar neurons, arbor
sizes agree with the rule. In Section \ref{sec:other} I consider
other factors which may affect arbor sizes.

\section{Topographic projection in one dimension}
\label{sec:1d}
Consider two parallel rows of evenly-spaced neurons, Figure~\ref{fig:cd},
with a topographic wiring diagram characterized by divergence, $D$, and
convergence, $C$. The goal is to find axonal and denritic arbor sizes
which minimize the combined length of axons and dendrites. I compare
different arbor arrangements by calculating wirelength per unit length
of the rows, $L$.  I assume that input/output rows are close to each
other and include in the calculation only those parts of the wiring
which are parallel to the neuronal rows.

I start by considering a special case where each input neuron
connects with only one output neuron ($D=1$), Figure~\ref{fig:menorah}a.
There are two limiting arrangements satisfying the wiring diagram: type I
has wide dendritic arbors and no axonal arbors,
Figure~\ref{fig:menorah}b; type II has wide axonal arbors and no
dendritic arbors, Figure~\ref{fig:menorah}c. Intuitively, the former
arrangement has smaller wirelength: short axons synapsing onto a
common bus-like dendrite is better than long axons from each input
neuron synapsing onto a short dendrite. To confirm this I calculate 
wirelength in the two extreme arrangements for $D=1$, see Methods. 
\begin{equation}
\label{li}
L_I=(1-1/C).
\end{equation}
\begin{equation}
\label{lii}
L_{II}=\left \{ \begin{array}{ll}   C/4,  
&  C-{\rm even,} \\ (C-1/C)/4, &  
C-{\rm odd.} \end{array} \right.
\end{equation}

These results show that for $D=1$ and $C\leq 3$ the two arrangements
have the same wirelength. For $D=1$ and $C>3$ the arrangement with
wide dendritic arbors and no axonal arbors (Type I) has smaller
wirelength than the arrangement with wide axonal arbors and no
dendritic arbors (Type II).

I can readily apply this result to another special case, $C=1$, by
invoking the symmetry of the problem in respect to the direction of
the signal propagation. I can interchange words ``axons'' and
``dendrites'' and variables $D$ and $C$ in the derivation and use the
above argument.  For $C=1$ and $D\leq 3$ the two extreme arrangements
have the same wirelength, while for $D>3$ the arrangement with wide
axonal arbors (Type II) has shorter wiring than the arrangement with
wide dendritic arbors (Type I).

Next, I consider the case when both convergence and divergence are
greater than one ($D,C>1$). For the two extreme arrangements I get (see Methods):
\begin{equation}
\label{dd}
L_{I}=D(1-1/C),
\end{equation}
\begin{equation}
\label{cc}
L_{II}=C(1-1/D).
\end{equation}
Comparison of the two expressions reveals the following. If divergence
is less than convergence then the optimal arrangement has wide
dendritic and no axonal arbors (Type I). If divergence is greater
than convergence then the optimal arrangement has wide axonal and no
dendritic arbors (Type II). If convergence and divergence are equal
both arrangements have the same wirelength.

I can restate this result by using the identity between the
divergence/convergence ratio and the neuronal density ratio (see
Methods): In the optimal arrangement the sparser layer has wide
arbors, while the denser layer has none.

So far I compared extreme arrangements with wide arbors in one row
and none in the other. What about intermediate arrangements, with both
axonal and dendritic arbors of non-zero width? To address this
question I consider the limit of large divergence and convergence 
factors ($C,D\gg 1)$. I find wirelength as a function of
the axonal arbor size $s_a$ (see Methods):
\begin{equation}
\label{cll}
L(s_a)=n_1s_a\left(1-\frac{D}{C}\right)+D.
\end{equation}
Because $0<s_a<C/n_1$ I find the following. If $D/C<1$ then the
minimal wirelength is achieved when $s_a=0$, arrangement with wide
dendritic and no axonal arbors (Type I). If $D/C>1$ then the minimal
wirelength is achieved when $s_a=C/n_1$, arrangement with wide axonal
and no dendritic arbors (Type II). If $D/C=1$ then all possible axonal
arbor widths give the same wirelength.

This proves that for $C,D\gg1$ extreme arrangements minimize
wirelength. In cases of small $C$ and $D$ I checked intermediate
solutions one by one. In many cases intermediate arrangements have the
same wirelength as the extreme solution. However, only for a few 
``degenerate'' $D,C$ pairs
there are equally good intermediate arrangements with the reverse ratio of 
average axonal and dendritic arbor sizes relative to the extreme solution.

\begin{figure}
\centerline{
\psfig{file=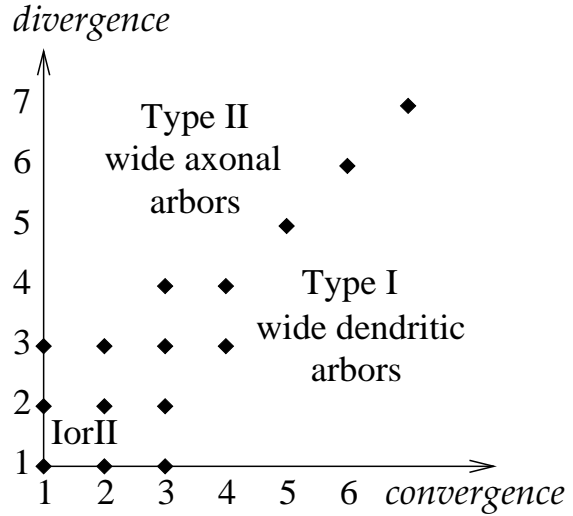,width=3.1in}
}
\vspace{0.1in} 
\setlength{\columnwidth}{3.4in}
\caption{Phase diagram for the one-dimensional projection. Optimal 
arrangements for possible pairs of convergence and divergence are shown.
Diamonds designate projections where wirelength is minimized by
both wide axonal and wide dendritic arbor arrangements.
\label{fig:phase}
}
\vspace{-0.1in}
\end{figure}

My results are conveniently summarized on the phase diagram,
Figure~\ref{fig:phase}, which shows optimal arrangements for various
pairs of divergence and convergence factors. I mark the
``degenerate'' $D,C$ pairs by diamonds on the phase diagram,
Figure~\ref{fig:phase}.

What if axons and dendrites have different crossectional areas? The
principle of wiring economy requires that wire volume rather than wire
length should be minimized. I can modify the arguments of this Section
by including the cross-sectional areas of the processes. I find for
$D,C\gg 1$ that if divergence/convergence ratio is less than the ratio
of axonal and dendritic cross-sections then the optimal arrangement
has wide dendritic and no axonal arbors (Type I). In the opposite case
I find wide axonal and no dendritic arbors (Type II).

\section{Topographic projection in two dimensions}
\label{sec:2d}

Consider two parallel layers of neurons with densities $n_1$ and
$n_2$. The topographic wiring diagram has divergence and convergence
factors, $D$ and $C$, requiring each input neuron to connect with $D$
nearest output neurons and each output neuron with $C$ nearest input
neurons.  Again, the problem is to find the arrangement of arbors
which minimizes the total length of axons and dendrites. For different
arrangements I compare the wirelength per unit area, $L$. I assume
that the two layers are close to each other and include only those
parts of the wiring which are parallel to the layers.

\begin{figure}
\centerline{
\psfig{file=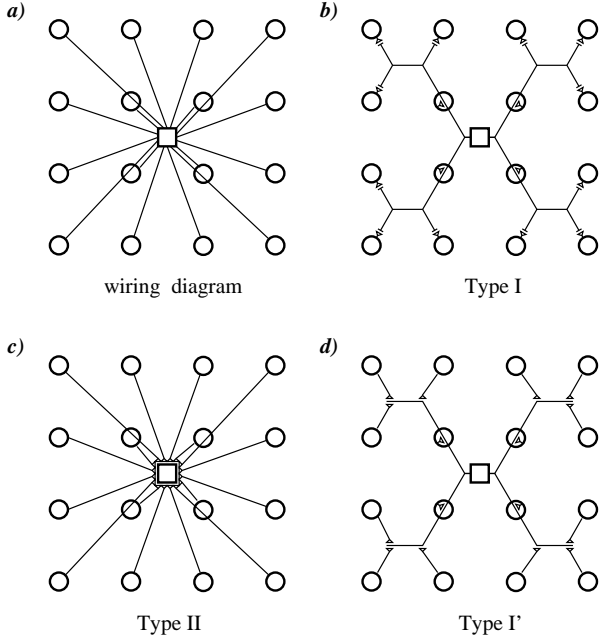,width=3.1in}
}
\vspace{0.1in} 
\setlength{\columnwidth}{3.4in}
\caption{
Different arrangements implement the same wiring diagram in two dimensions.
(a) Topographic wiring diagram with $D=1$ and $C=16$. (b) Arrangement with
wide dendritic arbors and no axonal arbors (Type I). (c) Arrangement with
wide axonal arbors and no dendritic arbors (Type II). Because convergence
exceeds divergence type I has shorter wiring than type II. (d) Intermediate
arrangement which has the same wiring length as type I.
\label{fig:por}
}
\vspace{-0.1in}
\end{figure}

I start with a special case where each input neuron connects with only
one output neuron ($D=1$). Consider an example with $C=16$ and neurons
arranged on a square grid in each layer, Figure~\ref{fig:por}a.  Two
extreme arrangements satisfy the wiring diagram: type I has wide
dendritic arbors and no axonal arbors, Figure~\ref{fig:por}b; type II
has wide axonal arbors and no dendritic arbors, Figure~\ref{fig:por}c. I
take the branching angles equal to $120^0$, an optimal value for
constant crossectional area.\cite{Chern} Assuming ``point'' neurons
the ratio of wirelength for type I and type II arrangements:
\begin{equation}
\frac{L_{I}}{L_{II}}\approx 0.57.
\end{equation}

Thus, the type I arrangement with wide dendritic arbors has shorter 
wire length. This conclusion holds for other convergence values much
greater than one, provided $D=1$. However,
there are other arrangements with non-zero axonal arbors that give the 
same wire length. One of them is shown in Figure~\ref{fig:por}d. Degenerate
arrangements have axonal arbor width $0<s_a<1/\sqrt{n_1}$ where the upper
bound is given by the approximate inter-neuronal distance. This means
that the optimal arbor size ratio for $D=1$
\begin{equation}
\label{sql}
\frac{s_d}{s_a}>\sqrt{\frac{n_1}{n_2}}
\end{equation}

By using the symmetry in respect to the direction of signal propagation I
adapt this result for the $C=1$ case. For $D>1$, arrangements with wide 
axonal arbors and narrow dendritic arbors ($0<s_d<1/\sqrt{n_2}$) 
have minimal wirelength. These arrangements have arbor size ratio
\begin{equation}
\frac{s_d}{s_a}<\sqrt{\frac{n_1}{n_2}}.
\end{equation}

Next, I consider the case when both divergence and convergence are 
greater than one. Due to complexity of the problem I study
the limit of large divergence and convergence ($D,C\gg 1$). I find
analytically the optimal layout which minimizes the total length of 
axons and dendrites. Unlike the one-dimensional projection, optimal 
sizes of both axons and dendrites turn out to be non-zero.

\begin{figure}
\centerline{
\psfig{file=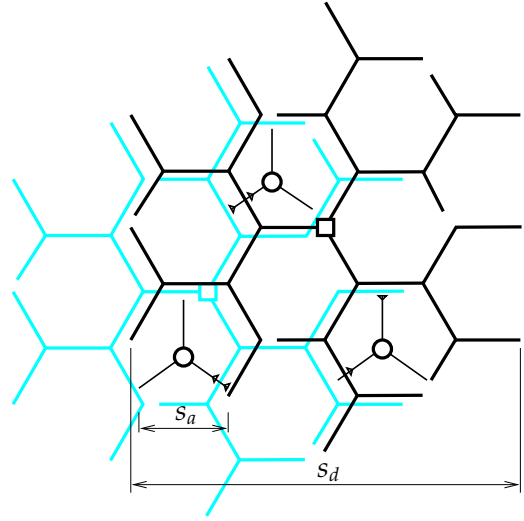,width=2.7in}
}
\vspace{0.1in} 
\setlength{\columnwidth}{3.4in}
\caption{Topographic projection between the layers of 
input (circles) and output (squares) neurons. For clarity, out of the 
many input and output neurons with overlapping arbors only few are shown.
The number of input neurons is greater than the number of output 
neurons ($C/D>1$). Input neurons have narrow axonal arbors of width $s_a$
connected to the wide but sparse dendritic arbors of width $s_d$. Sparseness
of the dendritic arbor is given by $s_a$ because all the input neurons
spanned by the dendritic arbor have to be connected. 
\label{fig:ar4}
}
\vspace{-0.1in}
\end{figure}

Notice that two neurons may form a synapse only if the axonal arbor of 
the input neuron overlaps with the dendritic arbor of the output
neuron in a two-dimensional projection, Figure~\ref{fig:ar4}. Thus the
goal is to design optimal dendritic and axonal arbors so that each
dendritic arbor intersects $C$ axonal arbors and each axonal arbor
intersects $D$ dendritic arbors.

To be specific, I consider a wiring diagram with convergence exceeding 
divergence, $C>D$ (the argument can be readily adapted for the opposite 
case). I make an assumption, to be verified later, that dendritic arbor
diameter $s_d$ is greater than axonal one, $s_a$. In this regime each
output neuron's dendritic arbor forms a sparse mesh covering the area
from which signals are collected, Figure~\ref{fig:ar4}. Each axonal
arbor in that area must intersect the dendritic arbor mesh to satisfy
the wiring diagram. This requires setting mesh size equal to the
axonal arbor diameter.

By using this requirement I express the total length of axonal and
dendritic arbors as a function of only the axonal arbor size, $s_a$.
Then I find the axonal arbor size which minimizes the total
wirelength.  Details of the calculation are given in Methods. 

Here, I give an intuitive argument for why in the optimal layout both
axonal and dendritic size are non-zero. Consider two extreme layouts.
In the first one, dendritic arbors have zero width, type II. In this 
arrangement axons have to reach out to every output neuron. For large
convergence, $C\gg 1$, this is a redundant arrangement because of the
many parallel axonal wires whose signals are eventually merged. In the
second layout, axonal arbors are absent and dendrites have to reach out
to every input neuron.  Again, because each input neuron connects to 
many output neurons (large divergence, $D\gg 1$) many dendrites run
in parallel inefficiently carrying the same signal. A non-zero
axonal arbor rectifies this inefficiency by carrying signals to several 
dendrites along one wire.

I find that the optimal ratio of dendritic and axonal arbor diameters
equals to the square root of the convergence/divergence
ratio, or, alternatively, to the square root of the neuronal density
ratio:
\begin{equation}
\label{res}
\frac{s_d}{s_a}=\sqrt{\frac{C}{D}}=\sqrt{\frac{n_1}{n_2}}
\end{equation}

Since I considered the case with $C>D$ this result also justifies
the assumption about axonal arbors being smaller than dendritic ones.

For arbitrary axonal and dendritic cross-sectional areas, $h_a$ and $h_d$,  
expressions of this Section are modified. The wiring economy principle
requires minimizing the total volume occupied by axons and dendrites 
resulting in the following relation for the optimal arrangement:
\begin{equation}
\label{res1}
\frac{s_d}{s_a}=\sqrt{\frac{Ch_a}{Dh_d}}=\sqrt{\frac{n_1h_a}{n_2h_d}}
\end{equation}
Notice that in the optimal arrangement the total axonal volume of input
neurons is equal to the total dendritic volume of the output neurons.
 
\section{Discussion}
\label{sec:disc}

\subsection{Comparison of the theory with anatomical data}
\label{sec:data}

This theory predicts a relationship between the con-/divergence ratio
and the sizes of axonal and dendritic arbors. I test these predictions
on several cases of topographic projection in two dimensions. The
predictions depend on whether divergence and convergence are both
greater than one or not. Therefore, I consider the two regimes
separately. 

First, I focus on topographic projections of retinal neurons whose
divergence factor is equal or close to one. Because retinal neurons use mostly
graded potentials the difference between axons and dendrites is small
and I assume that their cross-sectional areas are equal. The theory
predicts that the ratio of dendritic and axonal arbor sizes must be
greater than the square root of the input/output neuronal density
ratio, $s_d/s_a>(n_1/n_2)^{1/2}$ (Eq.\ref{sql}).

\begin{figure}
\centerline{
\psfig{file=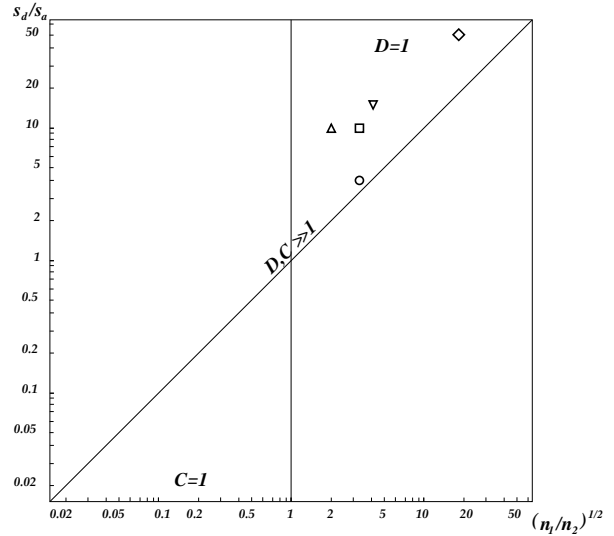,width=3.1in}
}
\vspace{0.1in} 
\setlength{\columnwidth}{3.4in}
\caption{Anatomical data for several pairs of retinal cell classes which form 
topographic projections with $D=1$.All the data points fall in the
triangle above the $s_d/s_a=(n_1/n_2)^{1/2}$ line in agreement 
with the theoretical prediction, Eq.\ref{sql}. The following data
has been used: $\circ$ - midget bipolar $\rightarrow$ midget 
ganglion (Watanabe \& Rodieck, 1989; Milam {\em et al.}, 1993; Dacey, 1993); 
$\sqcup$ - diffuse bipolar 
$\rightarrow$ parasol ganglion (Watanabe \& Rodieck, 1989;Grunert {\em et al.}, 1994) 
$\bigtriangledown$ -
rods $\rightarrow$ rod bipolar (Grunert \& Martin, 1991); $\bigtriangleup$ - cones 
$\rightarrow$ HI horizontals (Wassle {\em et al.}, 1989); $\diamond$ - rods 
$\rightarrow$ telodendritic arbors of HI horizontals (Rodieck, 1989).
\label{fig:data}
}
\vspace{-0.1in}
\end{figure}

I represent the data on the plot of the relative arbor diameter, $s_d/s_a$,
vs. the square root of the relative densities, $(n_1/n_2)^{1/2}$,
(Figure~\ref{fig:data}). Because neurons located in the same layer may
belong to different classes, each having different arbor size and 
connectivity, I plot data from different classes separately.
All the data points lie above the $s_d/s_a=(n_1/n_2)^{1/2}$ line in
agreement with the prediction.

Second, I apply the theory to cerebellar neurons whose divergence and 
convergence are both greater than one. I consider a projection from
granule cell axons (parallel fibers) onto Purkinje cells. 
Ratio of granule cells to Purkinje cells is 3300,\cite{Andersen}, 
indicating a high convergence/divergence ratio. This predicts a 
ratio of dendritic and axonal arbor sizes of 58. This is qualitatively
in agreement with wide dendritic arbors of Purkinje cells and no axonal
arbors on parallel fibers.

Quantitative comparison is complicated because the projection is not
strictly two-dimensional: Purkinje dendrites stacked next to each other
add up to a significant third dimension. Naively, given that the 
dendritic arbor size is about 400$\mu$m Eq.\ref{res} predicts axonal arbor
of about 7$\mu$m. This is close to the distance between two adjacent
Purkinje cell arbors of about 9$\mu$m. Because the length of parallel 
fibers is greater than 7$\mu$m absence of axonal arbors comes as no surprise.

In general, application of the rule requires some care because it was derived
for a simplified model. I considered a topographic projection only between  
a single pair of layers. However, neurons often make connections to different
layers. In particular, dendritic arbors of the output layer may be determined by 
connections other than to the input layer. 
For example, consider the topographic projection from thalamus to the primary
visual cortex. One may think that because the density of magnocellular
thalamic afferents is smaller than neurons in layer 4C$\alpha$ ($80mm^{-2}$
compared with $1.8\cdot 10^4mm^{-2}$)\cite{Peters} then the axonal arbors
should be wider than the dendritic ones. Although this is true 
($600\mu m$\cite{BL} compared with $200\mu m$\cite{WC}) the majority of inputs
to layer 4C$\alpha$ are intra-cortical.\cite{Peters} Therefore, the 
dendritic arbor size may be determined by these other projections.

\subsection{Other factors affecting arbor sizes}
\label{sec:other}

One may argue that dendrites and axons have functions other than
linking cell bodies to synapses and, therefore, the size of the arbors
may be dictated by other considerations. Although I can not rule out
this possibility, the {\it primary} function of axons and dendrites is
to connect cell bodies to synapses in order to conduct nerve pulses
between them. Indeed, if neurons were not connected more sophisticated
effects such as non-linear interactions between different dendritic
inputs could not take place. Hence the most basic parameters
of axonal and dendritic arbors such as their size should follow from
considerations of connectivity.

Another possibility is that the size of dendritic arbors is dictated by 
the surface area needed to arrange all the synapses. This argument does not 
specify the arbor size, however: a compact dendrite of elaborate shape 
can have the same surface area as a wide dendritic arbor.

Finally, agreement of the predictions with the existing anatomical data 
suggests that the rule is based on correct principles. Further extensive 
testing of the rule is desirable. Violation of the rule in some system 
would suggest the presence of other overriding considerations in the design 
of that system, which is also interesting.

In conclusion, I propose a rule relating connectivity of neurons to their
morphology based on the wiring economy principle. This rule may be used to 
infer connections between neurons from the sizes of their axonal and 
dendritic arbors.

\section{Methods}

I frequently use the following identity\cite{PurRub} relating 
convergence/divergence ratio and neuronal densities ratio:
\begin{equation}
\label{cnd}
\frac{C}{D}=\frac{n_{1}}{n_{2}}.
\end{equation}
To prove it, I calculate the number of connections (or synapses, if
connections are monosynaptic) per unit length in two ways. The 
number of connections (or synapses) is the number of input
neurons, $n_1$, times divergence, $D$. At the same time, the number
of connections (or synapses) is the number of output neurons, $n_2$,
times convergence, $C$. Since the answer should not depend on the
argument, $n_{1}D=n_{2}C$ and Eq.\ref{cnd} follows trivially.

\subsection{Projection in one dimension} 

First, consider the case of $D=1$. In type I arrangement (Figure
\ref{fig:menorah}b), the size of a
dendritic arbor, $s_d$, is the inter-neuronal spacing $1/n_1$ times the number 
of inter-neuronal intervals covered by the arbor, $C-1$:
\begin{equation}
s_d=(C-1)/n_1.
\end{equation}
The number of dendritic arbors per unit length is equal to the density
of output neurons $n_2$. The combined dendritic arbor length per unit
length is $n_2s_d$. Since the axonal arbors do not contribute, the
total wire length per unit length:
\begin{equation}
\label{lnn}
L_I=n_2s_d=n_2(C-1)/n_1.
\end{equation}
By using Eq.\ref{cnd} and recalling that $D=1$ I get Eq.\ref{li}.

In type II arrangement (Figure~\ref{fig:menorah}c), the wire length is
equal to the sum of the lengths of axons converging on each output
neuron multiplied by the neuronal density in the output layer $n_2$:
\begin{equation}
L_{II} = \left \{ \begin{array}{ll}  n_2((C-1)+(C-3)+...+1)/n_1=\\
\hspace{2cm} =n_2C^2/4n_1,  &  C-{\rm even,} \\ 
n_2((C-1)+(C-3)+...+0)/n_1=\\
\hspace{2cm} =n_2(C^2-1)/4n_1, &  C-{\rm odd.} \end{array} \right.
\end{equation}
Using Eq.\ref{cnd} I express the result in terms of convergence alone $(D=1)$
and get Eq.\ref{lii}.

Now consider the case of $D,C>1$. By using Eq.\ref{cnd} I find 
from Eq.\ref{lnn} that
\begin{equation}
L_{I}=D(1-1/C).
\end{equation}
This is Eq.\ref{dd} of the main text. By using the symmetry in respect to 
the direction of signal propagation I find Eq.\ref{cc} of the main text.

\begin{figure}
\centerline{
\psfig{file=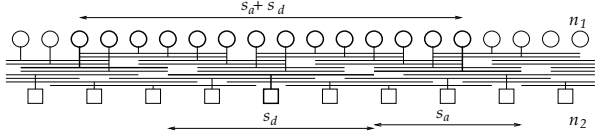,width=3.1in}
}
\vspace{0.1in} 
\setlength{\columnwidth}{3.4in}
\caption{
Topographic projection between the rows of input (circles) and
output (squares) neurons. The density of input neurons is greater than
the output, $C/D>1$. The sizes of the axonal arbors are $s_a$ and of
dendritic arbors, $s_d$. We show in bold the cell bodies of the
$C$ input neurons projecting to one output neuron (bold).
\label{fig:ar1}
}
\vspace{-0.1in}
\end{figure}

Next, I consider an arrangement with arbitrary sizes of axonal, $s_a$, and 
dendritic, $s_d$, arbors, Figure~\ref{fig:ar1}, in the limit of $D,C\gg 1$.  
To satisfy the wiring diagram each input neuron must connect with $D$ output
neurons and each output neuron must connect with $C$ input neurons. This 
places a constraint on the sum of axonal and dendritic arbor widths:
\begin{equation}
\label{ss}
s_a+s_d=D/n_2=C/n_1.
\end{equation}

Therefore, axonal arbor width can take values $0<s_a<C/n_1$.  The
total wirelength per unit length is:
\begin{equation}
L=s_an_1+s_dn_2.
\end{equation}
Using Eq.\ref{cnd},\ref{ss} I get Eq.\ref{cll} of the main text.

\subsection{Projection in two dimensions}

I consider the case of $C,D\gg 1$, Fig.\ref{fig:ar4}. The following 
calculation is valid to the leading order in $D$ and $C$. I omit
numerical factors of order one which depend on the precise geometry of
axonal and dendritic arbors.  The total length of a dendritic arbor,
$l_d$, is equal to the number of periods in the mesh $s_d^2/s_a^2$ times
the mesh size, $s_a$:
\begin{equation}
\label{sa}
l_d=\frac{s_d^2}{s_a}
\end{equation}

The size of the dendritic arbor, $s_d$ follows from expressing
convergence as the product of the area covered by the dendritic arbor
times the density of input neurons $C=s_d^2n_1$:
\begin{equation}
\label{la}
s_d^2=\frac{C}{n_1}.
\end{equation}
Substituting this into Eq.\ref{sa} I find:
\begin{equation}
l_d=\frac{C}{n_1s_a}
\end{equation}
The length of an axonal arbor is approximately given by its size:
\begin{equation}
l_a=s_a.
\end{equation}
Then the total wirelength per unit area is:
\begin{equation}
L=l_dn_2+l_an_1=\frac{Cn_2}{n_1s_a}+s_an_1.
\end{equation}

In order to find the optimal axonal arbor size $s_a$, I differentiate 
wirelength in respect to $s_a$ and set the derivative to zero.
\begin{equation}
\frac{\partial L}{\partial s_a}=-\frac{Cn_2}{n_1s_a^2}+n_1=0.
\end{equation}
Solution of this equation gives the optimal size of an axonal arbor,
$s_a$:
\begin{equation}
s_a=\sqrt{\frac{Cn_2}{n_1^2}}=\sqrt{\frac{D}{n_1}}
\end{equation}
By using Eq.\ref{la} I get the size of the dendritic arbor,
\begin{equation}
s_d=\sqrt{\frac{C}{n_1}}
\end{equation}
The last two equations combined give Eq.\ref{res} of the main text.

\section*{Acknowledgements}

I have benefited from helpful discussions with E.M. Callaway, E.J.
Chichilnisky, H.J. Karten, C.F. Stevens and T.J. Sejnowski. I am
grateful to A.A. Koulakov for making several valuable suggestions. I
thank G.D. Brown for suggesting that the size of axonal and dendritic
arbors may be related to con-/divergence.  This research was supported
by the Sloan Foundation.


\end{multicols}
\end{document}